\documentclass{iau}
\usepackage{graphicx}

 \title[Polyynyl-substituted PAH molecules] %% give here short title %%
{Polyynyl-substituted PAH molecules \\ and DIB carriers}

\author[Rouill{\'e} et al.]   %% give here short author list %%
{Ga{\"e}l Rouill{\'e}$^1$,
Cornelia J{\"a}ger$^1$,
Friedrich Huisken$^1$, \\
\and Thomas Henning$^2$}

\affiliation{$^1$Laboratory Astrophysics Group of the Max Planck Institute for Astronomy \\ at the Friedrich Schiller University Jena, Institute of Solid State Physics, \\ Helmholtzweg 3, 07743 Jena, Germany \\ email: {\tt cornelia.jaeger@uni-jena.de} \\[\affilskip]
$^2$Max Planck Institute for Astronomy, K{\"o}nigstuhl 17, 69117 Heidelberg, Germany}

\pubyear{2013}
\volume{297}  %% insert here IAU Symposium No.
\pagerange{1--5}
\jname{The Diffuse Interstellar Bands}
\editors{J. Cami \& N. Cox, eds.}
\begin{document}

\maketitle

\begin{abstract}
Polycyclic aromatic hydrocarbon (PAH) molecules have been long considered promising candidates for the carriers of the diffuse interstellar bands (DIBs). The PAH-DIB hypothesis, however, raises two major issues. First, the number of interstellar PAH species is potentially orders of magnitude larger than the number of DIBs. Second, the absorption spectrum of a PAH is in general dominated by bands found at UV wavelengths while, conversely, DIBs are absent from the UV wavelength domain and arise at visible and near IR wavelengths. These issues do not necessarily weaken the PAH-DIB hypothesis and can actually allow us to refine it. In that context, we analyze the UV/vis absorption spectra of PAH molecules isolated in Ne matrices and propose that polyynyl-substituted PAHs, or similar species, are valid candidates for the carriers of the DIBs. Finally, a possible lifecycle for DIB-carrying PAHs is presented.
\keywords{ISM: lines and bands, ISM: molecules, methods: laboratory, molecular data}
\end{abstract}

\firstsection % if your document starts with a section,

\section{Introduction}

In 1985, separate groups proposed polycyclic aromatic hydrocarbon (PAH) molecules as candidates for the carriers of the diffuse interstellar bands (DIBs), thereby introducing the PAH-DIB hypothesis (\cite[Crawford \etal\ 1985]{Crawford85}; \cite[L{\'e}ger \& d'Hendecourt 1985]{Leger85}; \cite[van der Zwet \& Allamandola 1985]{vanderZwet85}). An issue was immediately raised by \cite[L{\'e}ger \& d'Hendecourt (1985)]{Leger85} who noted that the number of  PAH species was extremely large compared to the number of DIBs. It was raised again later by \cite[Salama \etal\ (1996)]{Salama96} who estimated the number of strong-to-moderate DIBs to be about 50. For comparison, the number of PAH species comprising 12 fused six-membered carbon rings, hence containing at most 50 C atoms, is 683\,101 (\cite[Tan 2009]{Tan09}).

A second issue arises when comparing the UV/vis spectra of PAH molecules with the DIBs. To date, DIBs have been observed at visible and near IR (NIR) wavelengths, between 4000 and 20\,000 {\AA} (\cite[Joblin \etal\ 1990]{Joblin90}; \cite[Hobbs \etal\ 2008, 2009]{Hobbs08,Hobbs09}; \cite[Geballe \etal\ 2011]{Geballe11}). Searches for DIBs in the UV domain have not been positive yet (\cite[Clayton \etal\ 2003]{Clayton03}; \cite[Gredel \etal\ 2011]{Gredel11}; \cite[Salama \etal\ 2011]{Salama11}), indicating that only weak ones may ever be found at wavelengths shorter than 4000 {\AA}. By contrast, the first electronic transition of a PAH molecule, which may give rise to bands in the vis/NIR domain, is in general weaker than transitions of higher order that give strong absorption bands at UV wavelengths. Moreover, there is no firm evidence yet that any two DIBs share the same carrier, whereas PAH spectra often show several bands of comparable intensities.

Thus, assuming that at least some of the DIB carriers are actually PAHs, these PAHs must stand out among the others by their greater number and their UV/vis/NIR spectrum must be dominated by a single, strong band lying between 4000 and 20\,000 {\AA}. These premises may seem to weaken the PAH-DIB hypothesis since they imply that the most abundant interstellar PAHs have uncommon spectral properties. To the contrary, our recent spectroscopic study of polyynyl-substituted PAHs (\cite[Rouill{\'e} \etal\ 2012]{Rouille12}) allows us to propose, for interstellar PAHs, a lifecycle that is consistent with the PAH-DIB hypothesis.

\section{The UV/vis spectrum of polyynyl-substituted PAHs}

\begin{figure}[b]
% \vspace*{-2.0 cm}
\begin{center}
 \includegraphics[width=5.3in]{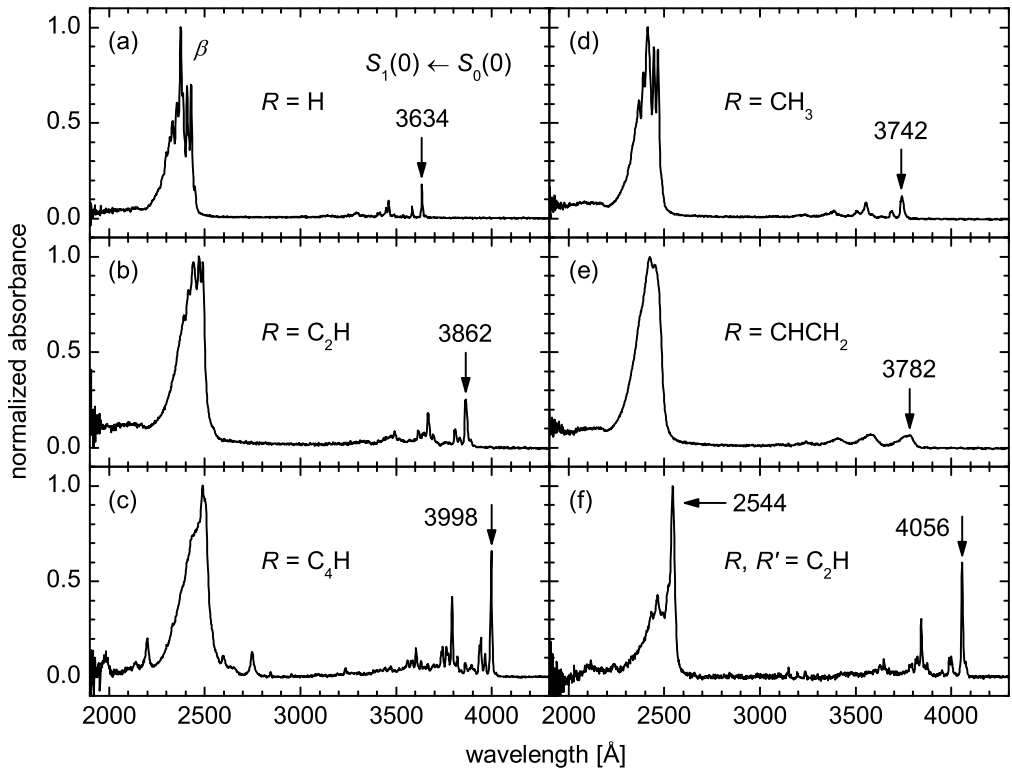}
% \vspace*{-1.0 cm}
 \caption{Absorption spectra of anthracene and $R$ derivatives isolated in Ne matrices at 6 K. (a) Anthracene. (b) 9-ethynylanthracene. (c) 9-butadiynylanthracene. (d) 9-methylanthracene. (e) 9-vinylanthracene. (f) 9,10-diethynylanthracene.}
\label{fig:Ant}
\end{center}
\end{figure}

We recently reported the UV/vis absorption spectrum of several polyynyl-substituted PAH species isolated in Ne matrix at 6 K (\cite[Rouill{\'e} \etal\ 2012]{Rouille12}). This study dealt with a selection of ethynyl and butadiynyl derivatives of anthracene, phenanthrene, and pyrene. Since then, we have measured the spectra of other derivatives of anthracene to obtain more information on the effect of various substitution schemes. These species were 9-methylanthracene, 9-vinylanthracene, and 9,10-diethynylanthracene. The apparatus for matrix isolation spectroscopy
% (MIS)
and the conditions of the first measurements were described in detail in \cite[Rouill{\'e} \etal\ (2012)]{Rouille12}. We have carried out the new measurements following the same protocol, applying specific conditions. For 9-methylanthracene and 9-vinylanthracene, the sample was kept at room temperature in the effusive source, the Ne gas flow rate was 5 sccm (standard cubic centimeter per minute), and the deposition lasted 30 min. The spectrum of 9-ethynylanthracene has been newly measured under these conditions. For 9,10-diethynylanthracene, the sample was brought to 56 $^{\circ}$C, increasing to 65 $^{\circ}$C during a deposition that lasted 60 min with a Ne gas flow rate of 5 sccm.

The spectra of the polyynyl-substituted PAHs have already been discussed (\cite[Rouill{\'e} \etal\ 2012]{Rouille12}). Briefly, we have found that, for a given PAH, the substitution of a H atom with an ethynyl chain shifts the absorption bands toward longer wavelengths and decreases the relative peak intensity of the generally prominent $\beta$-bands (\cite[Clar 1950]{Clar50}). The latter effect, accompanied by the broadening of the bands, has been attributed to fast deexcitation through internal energy conversion. Upon substitution with a butadiynyl chain, the effects are stronger. Thus, for instance, while the spectrum of phenanthrene is dominated by the $\beta$-band at 2428 {\AA}, the spectrum of 9-butadiynylphenanthrene is dominated by the origin band of the $S_2 \leftarrow S_0$ transition at 3156 {\AA}. 

Figure~\ref{fig:Ant} illustrates the effect of different substitutions on the spectrum of anthracene. Background correction has been carried out for the 9-butadiynyl and 9,10-diethynyl derivatives as the long depositions produced thick matrices that caused significant scattering. When comparing the spectra of the 9-ethynyl and 9-butadiynyl derivatives of anthracene [Fig.\,\ref{fig:Ant}(b--c)] with that of the unsubstituted molecule [Fig.\,\ref{fig:Ant}(a)], one observes the effects described in the previous paragraph, even though they are not as spectacular as in the spectra of the phenanthrenes and pyrenes (\cite[Rouill{\'e} \etal\ 2012]{Rouille12}). Substitution with a methyl or a vinyl group [Fig.\,\ref{fig:Ant}(d--e)] also causes redshifts, which were measured earlier in supersonic jets for the $S_1 \leftarrow S_0$ transition (\cite[Syage \etal\ 1985]{Syage85}; \cite[Werst \etal\ 1987]{Werst87}). Our spectra show that these shifts are not as large as those observed upon substitution with an ethynyl or a butadiynyl group. They also show that substitution with a methyl or a vinyl group does not affect the relative peak intensities to the advantage of the bands of the $S_1 \leftarrow S_0$ transition. These bands appear to be broadened in the spectrum of the 9-methyl and 9-vinyl derivatives. Comparison with the jet-cooled spectra, in which the bands are narrow, leads us to attribute these broadenings to a site effect caused by the nonplanar structure of the species, with, in addition, the overlap of numerous bands in the case of 9-vinylanthracene. Of course these effects also affect the $\beta$-band in a corresponding proportion. Finally, the spectrum of 9,10-diethynylanthracene reveals the largest redshifts [Fig.\,\ref{fig:Ant}(f)]. Even though the bands of the $S_1 \leftarrow S_0$ transition are strong, the spectrum is dominated by a relatively narrow component of the $\beta$-band, possibly its origin band. This feature may be attributed to the particular substitution scheme, i.e., chains attached at opposite carbons of an aromatic ring.

\section{Solving the issues in the PAH-DIB hypothesis}

The simplest way to solve the two issues in the PAH-DIB hypothesis is to consider them as interdependent. In other words, the fact that DIB-carrying PAHs have larger populations than the other PAHs must be related to the fact that they do not have UV transitions with high peak intensities. Obviously, the same principle can be generalized to DIB carriers that are not PAHs.

To explain the larger population of DIB-carrying PAHs, \cite[L{\'e}ger \& d'Hendecourt (1985)]{Leger85} suggested the effect of a selection mechanism. They took a higher thermodynamic stability as an example of selection criterion, which would favor populations of fully-benzenoid PAHs such as hexa-peri-hexabenzocoronene. Note that an attempt to detect this species in several astronomical objects was unsuccessful (\cite[Gredel \etal\ 2011]{Gredel11}). Fundamentally, however, the selection results from the balance between the formation and destruction processes of the molecules (\cite[Duley 2006]{Duley06}). Assuming that the DIB carriers are formed in the ISM, the formation process in the case of PAHs can be built on the ethynyl addition mechanism (\cite[Mebel \etal\ 2008]{Mebel08}). As for the main destruction process, the proposed interdependence with the electronic spectrum points toward a photophysical effect, i.e., photodissociation.

Let us consider that the selection criterion for DIB carriers is a higher photostability under exposure to the interstellar UV flux. Photodissociation is caused by the accumulation of a large enough amount of energy in a bond, or a small number of bonds, following the absorption of a photon. It can be prevented by fast intramolecular vibrational energy redistribution (IVR), followed by, e.g., internal conversion and photoemission (\cite[Allamandola \etal\ 1989]{Allamandola89}). The potential of a molecule for fast IVR and deexcitation can be indicated by the magnitude of the broadening of the absorption features in $S_n \leftarrow S_0$ transitions with $n$ $>$ 1 or 2, which belong generally to the UV wavelength domain. A strong broadening and the incidental lowering of the peak intensity of these features would make them extremely difficult to detect. Thus, we have found that a larger population can be related to an absorption spectrum characterized by bands with low peak intensities in the UV domain.

For PAHs, a higher photostability through fast IVR can be linked to the presence of side groups. Among the rare studies on the photostability of PAHs, one has shown that the presence of a side group decreases the photostability of neutral PAHs (\cite[Jochims \etal\ 1999]{Jochims99}). Note that while the groups selected for this study included the methyl and vinyl groups, it did not comprise polyynyl chains. Another study has reported the observation of stable PAH cations that carried either a CHCH chain or two CH groups (\cite[Ekern \etal\ 1998]{Ekern98}). Their photostability has been attributed to the presence of the side groups, possibly the CHCH chain, which would act as facilitator for IVR.

In the face of the spectra of polyynyl-substituted PAHs, we have already proposed that the broadening of the $\beta$-bands and the accompanying lowering of peak intensity are caused by fast internal conversion (\cite[Rouill{\'e} \etal\ 2012]{Rouille12}). We add now IVR as a contributor to a fast deexcitation process made possible by the presence of the polyynyl side chains. We expect polyynyl-substituted PAHs to be more photostable than unsubstituted PAHs and consequently than PAHs carrying other hydrocarbon groups (\cite[Jochims \etal\ 1999]{Jochims99}), in consistency with the spectra of Figure~\ref{fig:Ant}. We note that the deexcitation mechanism is facilitated with various degrees, being more efficient in phenanthrenes and pyrenes than in anthracenes. Moreover, following our line of thought, the spectrum of 9,10-diethynylanthracene, with a sharpened $\beta$-band, suggests that some substitution schemes with polyynyl groups can on the contrary lead to lower photostabilities.

After attributing to DIB-carrying PAH molecules a higher photostability arising from the presence of side groups, it remains to determine their formation process. They could be generated by the photodissociation of larger species as observed by \cite[Ekern \etal\ (1998)]{Ekern98} in their experiment. Instead of being a top-down process, however, the formation could proceed in a bottom-up fashion, taking advantage of the ethynyl addition mechanism (\cite[Mebel \etal\ 2008]{Mebel08}). This mechanism can take place in the ISM, for it is barrierless. The internal excess energy resulting from the addition of an ethynyl group can be released through the channel that also prevents photodissociation, that is fast IVR followed by internal conversion and photoemission.

Ultimately, DIB-carrying PAHs are destroyed in collisions with high-energy photons and particles constituting cosmic rays or, indirectly, by losing their photostability due to a structural transformation, e.g., after a reaction with another molecular species. This case can be illustrated using the spectra of anthracene, 9-ethynylanthracene, and 9,10-diethynylanthracene [Fig.\,\ref{fig:Ant}(a--b, f)]. While the addition of a C$_2$H group to anthracene can lead to the more photostable 9-ethynyl derivative, the further addition of a similar side chain may give 9,10-diethynylanthracene, which we suggest is less photostable. The fragments resulting from the destruction of the DIB carriers return to the ISM where they become available for new chemical processes, including the formation of new DIB carriers. Thus, we have established a lifecycle for DIB-carrying PAHs.

\section{Conclusion}

After comparing the UV/vis spectra of various PAHs isolated in Ne matrices at 6~K, we propose a single solution to two major issues of  the PAH-DIB hypothesis. This solution takes the form of substituted PAHs made particularly photostable by an efficient deexcitation mechanism that starts with IVR. More laboratory studies on the formation and destruction of PAHs are needed to confirm this conclusion. They should focus on substituted PAHs, more particularly on their formation at low temperature and their photostability.

\acknowledgment
The authors acknowledge the financial support of the Deutsche Forschungsgemeinschaft through project No. HU 474/24. They are grateful to H.-J. Kn{\"o}lker (Technical University Dresden) and coworkers who synthesized several of the substances required to carry out this study. They also thank Y. Carpentier who measured the spectrum of anthracene in Ne and M. Steglich for his assistance during experiments.

\end{document}